\documentclass[sigconf,natbib=true]{acmart}

\usepackage{pgfplots}
\usepackage{pgfplotstable}
\usepgfplotslibrary{groupplots}
\usepgfplotslibrary{statistics}
\AtBeginDocument{%
  }

\setcopyright{acmlicensed}
\copyrightyear{2025}
\acmYear{2025}
\acmDOI{XXXXXXX.XXXXXXX}

\usepackage{tikz}

\acmJournal{JACM}
\acmVolume{37}
\acmNumber{4}
\acmArticle{111}
\acmMonth{1}

\copyrightyear{2025}
\acmYear{2025}
\setcopyright{rightsretained}
\acmConference[CHI EA '25]{Extended Abstracts of the CHI Conference on Human Factors in Computing Systems}{April 26-May 1, 2025}{Yokohama, Japan}
\acmBooktitle{Extended Abstracts of the CHI Conference on Human Factors in Computing Systems (CHI EA '25), April 26-May 1, 2025, Yokohama, Japan}\acmDOI{10.1145/3706599.3720136}
\acmISBN{979-8-4007-1395-8/2025/04}

\begin{document}

\title{Exploring Culturally Informed AI Assistants: A Comparative Study of ChatBlackGPT and ChatGPT}

\author{Lisa Egede}
\authornote{Both authors contributed equally to this research.}
\affiliation{
\institution{Carnegie Mellon University}
  \city{Pittsburgh}
  \state{Pennsylvania}
  \country{USA}}
\email{lisaegede@cmu.edu}

\author{Ebtesam Al Haque}
\authornotemark[1]
\affiliation{
\institution{George Mason University}
  \city{Fairfax}
  \state{Virginia}
  \country{USA}}
\email{ehaque4@gmu.edu}

\author{Gabriella Thompson}
\orcid{0009-0008-7410-2591}
\affiliation{%
  \institution{The University of Texas at Austin}
  \city{Austin}
  \state{Texas}
  \country{USA}}
\email{gabriella.thompson@utexas.edu}

\author{Alicia Boyd}
\affiliation{%
  \institution{Yale University}
  \city{New Haven}
  \state{Connecticut}
  \country{USA}}
\email{alicia.boyd@yale.edu}

\author{Angela D. R. Smith}
\affiliation{%
  \institution{The University of Texas at Austin}
  \city{Austin}
    \state{Texas}
  \country{USA}}
\email{adrsmith@utexas.edu}
\orcid{0000-0001-5546-5452}

\author{Brittany Johnson}
\affiliation{%
  \institution{George Mason University}
  \city{Fairfax}
  \state{Virginia}
  \country{USA}}
\email{johnsonb@gmu.edu}

 \renewcommand{\shortauthors}{Egede et al.}

\begin{abstract}
In recent years, we have seen an influx in reliance on AI assistants for information seeking. Given this widespread use and the known challenges AI poses for Black users, recent efforts have emerged to identify key considerations needed to provide meaningful support. One notable effort is the development of ChatBlackGPT, a \textit{culturally informed} AI assistant designed to provide culturally relevant responses. Despite the existence of ChatBlackGPT, there is no research on when and how Black communities might engage with \textit{culturally informed} AI assistants and the distinctions between engagement with general purpose tools like ChatGPT.
To fill this gap, we propose a research agenda grounded in results from a preliminary comparative analysis of outputs provided by ChatGPT and ChatBlackGPT for travel-related inquiries. Our efforts thus far emphasize the need to consider Black communities' values, perceptions, and experiences when designing AI assistants that acknowledge the Black lived experience.

\end{abstract}




\maketitle

\section{Introduction}

AI assistants have become a widely relied-on resource, with people using them for information seeking, support, and simple task completions \cite{amer2024end}. 
AI assistants vary in modality, with text-based tools (\textit{e.g.,} OpenAI's ChatGPT, Google's Gemini, \textit{etc.}) and voice-based assistants being the most prominent (\textit{e.g.,} Alexa from  Amazon, Siri from Apple) \cite{rane2024gemini}. 
Users' modality choice is influenced by personal preferences and access to AI assistants~\cite{golden2018differences, mullen2024m}. Text-based assistants are particularly popular, and can be viewed as a ``stand in'' for search engines due to their ease of use, access, and reliability as an information-seeking resource ~\cite{xu2023chatgpt, jung2024we, walker2024they, bansal2024transforming}. 

For individuals who do have access to AI assistants, a deciding factor for using tools like ChatGPT is knowing whether these tools will work for them, which serves as a particular point of contention for Black users who have historically been mis- or under-represented in AI discourse~\cite{ngueajio2022hey, cunningham2024understanding, weidinger2021ethical}. In fact, prior research has emphasized the likelihood that Black users will have a negative experience as a result of inaccurate information or overt failure to understand or contextualize the culturally nuanced nature of their inquiries~\cite{klassen2024black, harrington2022s, brewer2023envisioning, mengesha2021don}. 


In this modern technological landscape, online communities and spaces that center on Black experiences have served as a powerful way to access culturally tailored information, seek community, and share resources \cite{klassen2021more, smith2024governance, egede2024us}. One of the most prevalent examples in the Black community is that of information seeking when traveling {~\cite{dillette2019tweeting}. Research suggests that Black users seeking travel-related inquiries leverage online spaces, where they can access insights from other Black travelers that may address concerns or perspectives overlooked by others~\cite{dillette2019tweeting, loewen2009sundown, peters2021instagramming, sutherland2019social}.
For example, in the U.S., there is a long history of ``sundown towns,'' which refers to a town that has historically excluded racial minorities and often enforces rules or threats requiring them to leave by sunset \cite{loewen2009sundown}. This points to an important consideration for Black travelers interested in using AI assistants for travel information -- accurately tailored information can ensure safety and peace of mind; inaccurate or misaligned information can lead to unsafe or even detrimental outcomes \cite{benjamin2024black}.





HCI research has highlighted the extent to which the Black lived experience is historically understudied across the design of AI systems, which has influenced how they are designed, and ultimately impacted how useful such tools are for Black users~\cite{egede2024us, ogbonnaya2020critical, slota2021methods, benjamin2019race, broussard2018artificial, gray_intersectional_2020}. The growth in dependence on AI assistants to serve a variety of needs for Black communities is prevalent in healthcare domains \cite{kim2022designing}, but minimal work or understanding of AI assistant use for broader culturally specific needs exist. While online resources for Black people are present, traditional AI assistants are currently not relied on as a resource to accurately capture the lived Black experience~\cite{slota2021methods, egede2024us}.


This raises an important question in the age of AI assistants: \textit{What does it look like to meaningfully design a culturally informed AI assistant?} Prior work suggests that the design and development of culturally informed AI assistant tools for Black communities require some level of lived experience~\cite{bray2021speculative,egede2024us, cunningham2023grounds}. However, in a field dominated by cisgender white males,~\footnote{\url{https://survey.stackoverflow.co/2022/\#developer-profile-demographics}} it is important we understand the necessary foundations for designing AI assistants that are capable of considering the perspective of historically marginalized users.


Towards this end, we propose a research agenda to explore how and in what ways culturally informed AI assistants provide support to Black users and how they compare to general purpose AI assistants~\cite{slota2021methods}.
To motivate and build a foundation for our research, we conducted a comparative analysis of ChatGPT -- a popular and widely used text-based AI assistant~\cite{chatgpt} -- and ChatBlackGPT -- a culturally informed AI assistant designed for Black communities by Black technologists~\cite{chatblackgpt2025, blackgpt_techcrunch2024}. We centered our preliminary efforts on travel-related inquiries, where we provided prompts to both ChatGPT and ChatBlackGPT from an existing dataset of inquiries from TikTok and Reddit curated specifically for creating culturally aware language technologies.~\footnote{\url{https://culturebank.github.io/}}


The goal of our preliminary experiment is to answer the following research questions:

\begin{description}
    \item[\textbf{RQ1}]  What are the similarities between ChatGPT and ChatBlackGPT when answering inquiries related to Black culture and travel?
    \item[\textbf{RQ2}] What are the distinguishing features of ChatBlackGPT when answering inquiries related to Black culture and travel?
\end{description}


By answering these research questions, we aim to ground our understanding of the key differences culturally tailored AI assistants have in response to culturally nuanced travel inquiries, as well as to emphasize the critical role culturally tailored AI assistants can play for Black users. This research agenda provides both CHI and the broader HCI community with foundational work to further understand how Black users perceive and engage with culturally tailored AI assistants designed by and for Black communities.

\section{Related Works}

\subsection{AI Assistants: Access, Trust, and Influences on User Perceptions}

Users who have historically opted for traditional search engines (\textit{e.g.,} Google) are leaning on AI assistants to support various tasks~\cite{lopez2018alexa}. AI assistant tools like ChatGPT, Gemini, and Siri are widely used to support task completion \cite{mun2024study}, as seen by users opting to use Amazon's Alexa to play music or for text-based tools like ChatGPT to craft personalized meeting agendas~\cite{lopez2018alexa}. The use cases for AI assistants have continued to expand, with the reliance on these tools as an information-seeking resource, with students in educational settings~\cite{adarkwah2023prediction}, and healthcare sectors \cite{zhan2024healthcare}. 

While the use of AI assistants are prominent, not everyone has equal access to its benefits. Work by Park \textit{et al.} outlines inequalities in access to digital assistants and found there to be perceptual divides around income, race, and gender, with the access-use barrier worsening preexisting offline socio-demographic inequalities \cite{park2022digital}. The technological gap is evident in how Black communities perceive AI assistants, along with the ability to which they can evaluate, participate, and critically engage with these tools \cite{park2022digital}. As a result, racially minoritzed communities experience disproportionate access to AI assistants, limiting exposure and feelings of autonomy in their engagement with tools \cite{field2021survey, wenzel2023can, ngueajio2022hey, mengesha2021don}.



The competence and overall usability of AI assistants are major components of the trust Black users may have in these systems, along with expectations that these tools will fail to meet their needs \cite{manzini2024should, chan2023harms}, given widely controversial failures (\textit{e.g.} Microsoft's AI chatbot Tay generating racist and sexist language \cite{wolf2017we}). Frameworks oriented around AI fairness highlight potential approaches to improve user trust in HCI research \cite{barocas2017bias, madaio2022assessing, raji2020closing}, with participatory and community-based design cited as powerful approaches to ensure AI systems are meeting the cultural needs of their users, by challenging power dynamics that often exist between researchers and designers versus communities and users \cite{harrington2019deconstructing, ogbonnaya2020critical, 10.1145/3240925.3240972}. 
While there has been attention focused on how researchers can approach engaging and co-designing with minoritized communities in the design of AI assistants, there is limited work on how culturally tailored AI assistants are being designed for Black communities. It can be inferred that this lack of intentional inclusion in AI systems has influenced Black communities to cultivate their own spaces and create tools that better center their needs ~\cite{egede2024us,benjamin2024black}.




\subsection{Blackness and AI Assistants: Designing Culturally Tailored Tools for Black Culture and Travel Information Seeking} 


Prior work has highlighted contrasts in Black users' experiences with AI assistants compared to other demographic groups \cite{cunningham2024understanding, brewer2023envisioning}, underscoring existing inequities and varying qualities of service \cite{mengesha2021don, beattie2022measuring}. Because AI assistants have difficulty understanding African-American Vernacular English (AAVE) or culturally nuanced prompts \cite{mengesha2021don, cunningham2024understanding, beattie2022measuring, park2022digital}, Black users tend to anticipate inaccurate responses, an interaction that reinforces linguistic discrimination and contributes to barriers of access \cite{schlesinger2018let}. Work by Appignani \textit{et al.} found that Microsoft Bing’s AI chatbot can engage in subtle forms of racism, such as tone policing, where legitimate user inquiries about racial themes are met with accusations of offensiveness or aggression \cite{appignani2024ai}. Similarly, Chandler \textit{et al.}'s work around the use of AI in Black women’s HIV self-education found notable differences in ChatGPT’s tone when responding to users based on race \cite{chan2023harms}. Anticipating microaggressive experiences while using AI assistants can amplify feelings of exclusion~\cite{wenzel2023can}.

In response, minoritized communities have begun to cultivate spaces as a way to center their lived experiences as seen by communities who identify as LGBTQIA+ \cite{jia2021needs, gray2009out, taylor2024mitigating}, communities with disabilities \cite{hofmann2020living}, and elder adults \cite{wu2012tangible, harrington2022s}. Black communities in particular have used digital spaces to share knowledge and culture, notably Black Twitter \cite{klassen2021more}, and BlackPlanet.com \cite{steele2023wish, byrne2007public}. While there has been a growth in centering culturally-tailored AI systems \cite{slota2021methods, mccall2021development}, Black communities are missing out on the rich experiences in a vast knowledge landscape. AI assistants \textit{explicitly} designed to center Black experiences remains limited \cite{slota2021methods}, with existing tools often constrained to certain domains, such as healthcare for elder adults \cite{harrington2022s, kim2022designing, chandler2024did} or education \cite{ogbo2024using}. This narrow scope highlights the need for culturally-tailored AI assistants that capture multifaceted aspects of Black culture and experiences.

While `For us, by us' -- the design philosophy emphasizing systems that are intended for a group should be developed by that very group, stemmed (academically) from disability studies, these efforts are beginning to emerge as they relate to race and the design of AI assistants~\cite{egede2024us, spiel2020nothing, traina2015participatory, bennett2018interdependence, gadiraju2023wouldn}. For example, tools like ChatBlackGPT illustrate the potential of intentional design approaches \cite{blackgpt_techcrunch2024}. Such systems are not only symbolically important but can also extend the life of curated community spaces, offering continuity even as platforms that once supported these spaces destabilize or cease to exist altogether (\textit{e.g.,} the ``dismantling'' of Black Twitter following Twitter's acquisition in 2022 \cite{nbc_black_twitter}). The existence of these tools alone emphasizes the cultural importance they may hold for Black users seeking support on experiences that require additional nuance, such as cultural experiences or travel-oriented topics.


Traveling while Black is a long documented issue, with prior work highlighting racist and microaggressive experiences they often encounter~\cite{peters2021instagramming, loewen2009sundown, benjamin2024black}. This has led to the creation of Black-centered travel groups like ``Outdoor Afro,'' as well as social media trends like \#TravelingWhileBlack~\cite{park2022flourishing}. Because Black travelers frequently face concerns that extend beyond the typical logistical questions traditional AI assistants are designed to answer, culturally tailored AI assistants can provide critical support in these scenarios. Specifically, a Black-developed and designed tool that is oriented around capturing nuanced questions, can be a reliable resource to utilize. The question remains: How accurate is a culturally tailored tool in capturing such nuanced prompts? Understanding how culturally tailored outputs differ from traditional AI assistants can highlight what users might need out of these tools.

\section{Methods}

To answer our research questions, we adopted a mixed methods approach. We first conducted a comparative analysis between ChatBlackGPT \cite{chatblackgpt2025} and ChatGPT \cite{chatgpt} using evaluation questions from CultureBank \cite{shi2024culturebank} combined with multiple text analysis metrics. 
Our analysis examined differences in response characteristics and linguistic patterns between the two AI assistants.
We then conducted a thematic analysis using \cite{clarke2017thematic}. The following sections provide details on our study design.
Given the exploratory nature of this work, we designed our comparative analysis to leverage commonly used methods and metrics for analyzing and interpreting textual data.

\subsection{Featured AI Assistants}
For this study, we leveraged two assistants: 1) ChatGPT and 2) ChatBlackGPT. ChatGPT launched November 30th of 2022, ChatGPT is a general-purpose generative AI tool founded by OpenAI~\cite{chatgpt}. It serves as a prominent information-seeking platform, with its primary goal being to generate natural language responses based on provided prompts. ChatGPT currently has 100+ million users, and the website sees nearly 1.5 billion visitors per month~\cite{Mahajan_2024a}. ChatBlackGPT launched June 19th of 2024, ChatBlackGPT is a culturally-informed generative AI tool that is specifically designed to provide inclusive content, with explicit consideration for topics related to the Black community and African diaspora~\cite{chatblackgpt2025,blackgpt_techcrunch2024}. 
ChatBlackGPT is a tool designed and developed \textit{with}, \textit{by}, and \textit{for} Black people.
Similar to ChatGPT, it generates responses based on text prompts.
ChatBlackGPT distinguishes itself from technologies like ChatGPT by emphasizing goals such as inclusive policies and empowerment through information.

 
\subsection{Dataset \& Prompts}

To curate responses for our comparative and qualitative analysis, we extracted data from CultureBank \cite{shi2024culturebank}, a dataset compiled from two major social media platforms: Reddit (\href{http://reddit.com}{Reddit.com}) and TikTok (\href{http://tiktok.com}{TikTok.com}). 
The dataset contains cultural descriptors sourced from Reddit (11K entries) and TikTok (12K entries), each with associated evaluation questions designed to probe cultural understanding. 
Topic analysis of these cultural descriptors revealed \textit{Community and Identity} as the predominant theme, with \textit{Communication and Language}, \textit{Social Norms and Values}, and \textit{Cultural Traditions and Festivals} also heavily featured. 
The Reddit-sourced portion showed particular emphasis on community-centric discussions, with "Community and Identity" and "Cultural Exchange" representing the majority of entries. 
For our specific analysis, we filtered the dataset to focus on entries related to the Black and African diaspora, resulting in 130 evaluation questions from the Reddit portion and 141 questions from the TikTok portion of CultureBank.
We used these questions as prompts for both ChatBlackGPT and ChatGPT, collecting responses from both tools. 

\subsection{Quantitative \& Qualitative Analysis}

\subsubsection{Sentiment Analysis}
We used the \textit{cardiffnlp/twitter-roberta-base-sentiment-latest} model ~\cite{loureiro-etal-2022-timelms}, a tool designed specifically for analyzing sentiment in conversational text. This model was trained on approximately 124 million tweets from January 2018 to December 2021 and subsequently fine-tuned for sentiment analysis using the TweetEval benchmark. The model classifies text into three sentiment categories — negative (0), neutral(1), and positive (2) — with each prediction including confidence scores for all three classes. 
This model was chosen because it works especially well with conversational language, making it ideal for analyzing AI assistant responses.

\subsubsection{Readability}
We adopted three widely-used metrics to evaluate the readability of responses: the Gunning Fog Index, Flesch Reading Ease Score, and the Flesch-Kincaid Grade Level. The \textbf{Gunning Fog Index} estimates the years of formal education needed to understand a text on first reading, using a formula that considers sentence length and the percentage of complex words (those with three or more syllables).  The \textbf{Flesch-Kincaid Grade Level} metric converts text complexity to U.S. grade levels using a formula that considers average word length (in syllables) and average sentence length (in words). The \textbf{Flesch Reading Ease Score} provides a numerical value between 0 and 100, with higher scores indicating more readable text. Scores of 90-100 are easily understood by an average fourth grader, 50-70 by high school students, and 0-30 by college graduates. 


\subsubsection{Thematic Analysis}
Considering prior research on bias in sentiment analysis tools \cite{diaz2018addressing} and to better ground our understanding of emerging themes in the outputs, we qualitatively coded responses from four sample prompts using an inductive coding approach ~\cite{clarke2017thematic}. 
Three of the researchers on our team coded all four sets of responses from ChatGPT and ChatBlackGPT, identifying similarities and differences in content and style. 
We aggregated the insights from each coder, merging any overlap in labeling and coming to agreement on all others. We acknowledge that qualitatively coding a larger sample could provide additional insights, and this represents a limitation of our approach. However, the purpose of this phase of analysis was exploratory. The findings from these coded responses provide a grounded starting point for understanding key themes, with the potential for future work to scale the analysis using larger data sets or complementary methods. 

\subsection{Statement of Positionality}
Studies have shown that researcher positionality can have a significant impact on the interpretation of data and experiences~\cite{patton2019annotating, cambo2022model}. This is especially the case when studying protected groups and historically marginalized communities. Our study aims to emphasize the value of culturally-informed AI assistants when supporting the lived Black experiences. As a team of Black and Brown women researchers, we bring relevant experiences and perspectives that support our ability to analyze responses from the Black perspective. As a team of college-educated researchers, we acknowledge that our collective research approaches and analysis were informed by our perspectives, education, and lived experiences within and outside of academia.

\section{Findings}




\subsection{Response Similarities (RQ1)}

In order to answer \textbf{RQ1} we present similarities found between the two AI assistants through a quantitative and qualitative analysis. Overall, our analysis found both AI assistants to exhibit similar tone and formatting structure. Additional examples of our analysis outputs can be found in the Github supplementary materials repository.\footnote{https://github.com/INSPIRED-GMU/chatgpt-chatblackgpt}

\subsubsection{Quantitative Analysis}


Our analysis showed that grade level distributions ranged from 8 to 14, with ChatBlackGPT responses scoring mostly at 12–14 (college level) and ChatGPT responses clustering at 8–10 (high school level) (Figure \ref{fig:readability}). ChatGPT was found to be slightly more readable -- perhaps due to list-style responses with less context (\textit{e.g.} fewer words). Our sentiment analysis found that both AI assistants responded to questions in a positive/neutral tone. Similarly, sentiment analysis of ChatBlackGPT's responses revealed a consistent pattern across all cultural topics: responses were uniformly neutral or positive in tone, with no negative sentiments detected. The majority of responses maintained a neutral sentiment, with positive sentiments appearing as a smaller but regular component across topics. This suggests the model may be calibrated to discuss cultural topics in a measured, constructive manner, avoiding negative characterizations. 

\begin{figure*}[htp]
    \centering
    \includegraphics[width=1\linewidth]{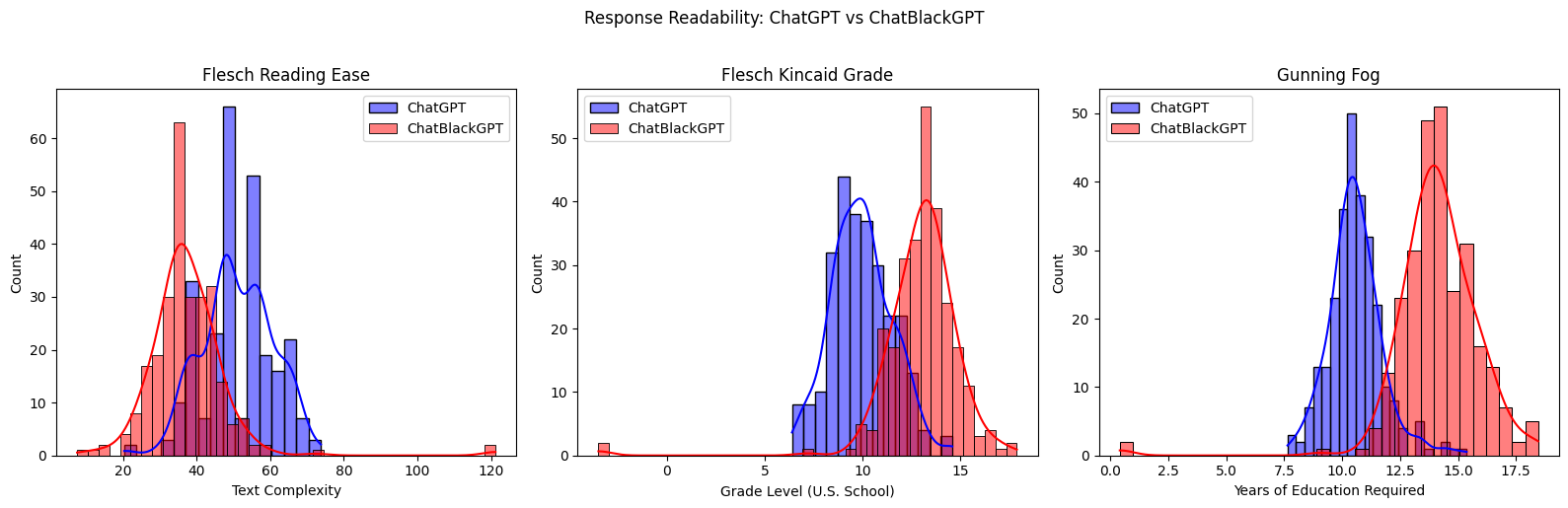}
    \caption{Readability of the Responses}
    \label{fig:readability}
\end{figure*}
    
\begin{figure*}[htp]
    \centering
    \includegraphics[width=1\textwidth]{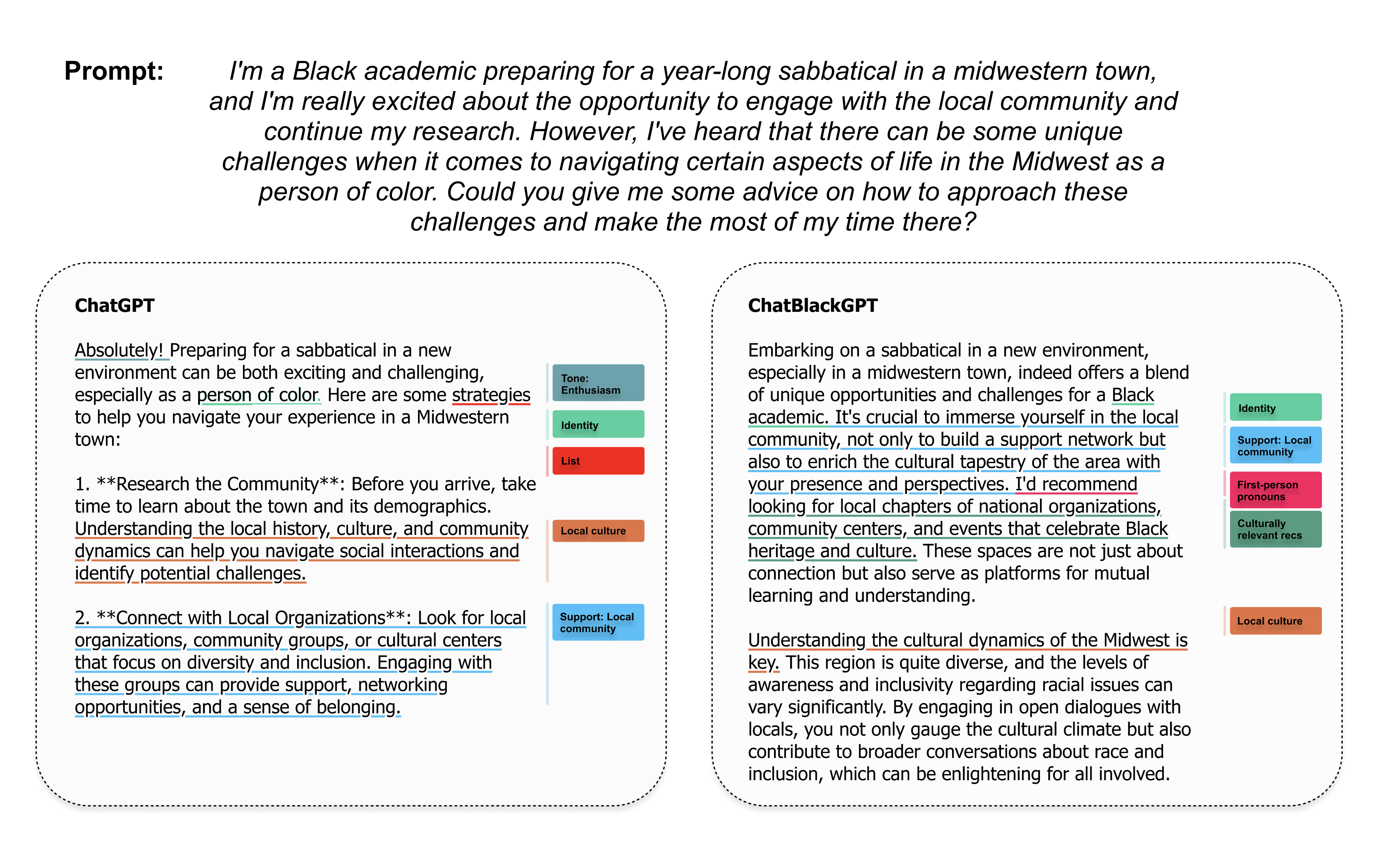}
    \caption{Shortened Sample Prompt and Output from ChatGPT and ChatBlackGPT}
    \label{fig:gptoutput}
\end{figure*}


\subsubsection{Qualitative Analysis}  
Our thematic analysis found both ChatGPT and ChatBlackGPT responded to questions with advice that modeled the same chronological structuring (before, during, and after the travel-related experience) (Figure \ref{fig:gptoutput}). While ChatGPT utilized bullet-style format, both AI assistants had a similar format structure of (opening, advice, and closing). There was observed overlap in the topic cover, and even across prompts. The most notable topics were advice regarding: researching and learning culture, being mindful and respectful in engagement with others, and documenting and sharing experiences throughout \cite{supplementary_chatblackgpt}.


\begin{figure*}[htp]
    \centering
    \includegraphics[width=1\textwidth]{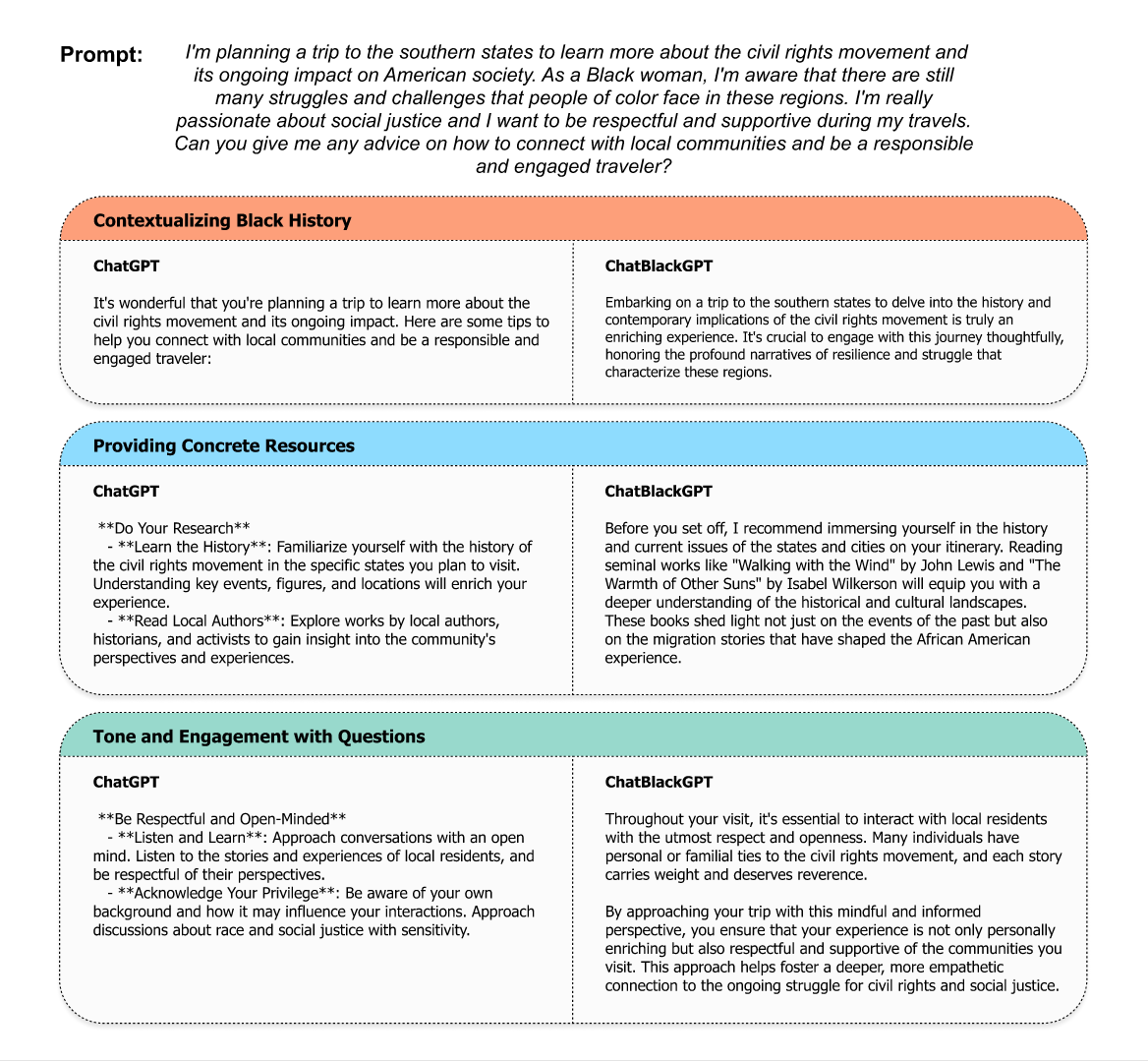}
    \caption{Distinguishing Features: Shortened Sample Prompt and Output}
    \label{fig:gptoutput2}
\end{figure*}

\subsection{Distinguishing Features (RQ2)}
A core goal for this work was to highlight distinguishable characteristics of culturally tailored AI assistant ChatBlackGPT. Our findings for \textbf{RQ1} illuminate quite a few similarities between ChatGPT and ChatBlackGPT on our dataset. To answer \textbf{RQ2} our qualitative analysis found a number of features in the responses provided by ChatBlackGPT that distinguishes it from ChatGPT when providing information for or about the Black community (Figure \ref{fig:gptoutput2}). 

\subsubsection{Contextualizing Black history}
When asked questions about specific locations or experiences, our analysis found that ChatBlackGPT responses went beyond giving high-level descriptions of specific locations, but historical context of how they came to be. For example, in response to a question asking about the Historically Black College/University (HBCU) experience, ChatBlackGPT details the history of HBCUs and the cultural significance they hold for Black communities. Providing concrete historical context to support suggestions on how to navigate these spaces emphasized the resource-oriented nature of their responses.

\subsubsection{Providing concrete resources}  
 ChatBlackGPT responses were often accompanied with culturally (and inquiry) relevant resources when applicable. Literature, locations, and other concrete resources were detailed, which was in contrast to ChatGPT's broad and non-direct responses. Additionally, ChatBlackGPT consistently suggested supporting Black-owned businesses -- a recommendation that they extended universally, including to non-Black users. ChatGPT made similar suggestions but only to Black users, (\textit{e.g.,} a Black woman asking for culturally tailored advice), rather than promoting this practice across all user interactions. 

\subsubsection{Tone and engagement with questions}
Tone was a notable theme, with ChatGPT adopting a more detached but enthusiastic response to culturally tailored questions. ChatBlackGPT adopted a more neutral conversational style-like tone, which was exemplified by its use of first-person pronouns. By using a personable and empathetic tone in their responses to culturally sensitive questions, there was an underlying perception that ChatBlackGPT was aware of the delicacy and sensitive nature of the questions being engaged with. For example, the highlighting of detailed safety concerns and considerations \cite{loewen2009sundown}, and explicit mentioning of the identity characteristics were noted in ChatBlackGPT's response. This was in contrast to ChatGPT, where observed questions specifically mentioning identity characteristics like ``Black woman'' were not explicitly mentioned in its response (\textit{e.g.,} the use of the words ``Black'' or ``woman''). ChatGPT's response to referring to Black communities and engaging with Black communities as ``engaging with the locals'' exhibited detached or avoidant language.

\section{Discussion and Future Work}

\subsection{Discussion of Preliminary Findings}
Between ChatGPT and ChatBlackGPT, communication style and overall tone of output responses stood out, with ChatGPT's overtly positive tone in response to sensitive topics being notable, especially when considering the history of ``tone policing'' of Black users in AI systems~\cite{appignani2024ai}. Such perceptions, combined with a tone that fails to align with the situation or the delicacy of the prompt, can influence how users' emotions and experiences are perceived and addressed by AI systems. Insight into how Black users perceive tone in AI assistants and the influence it has on their decision making skills remains to be explored. These interactions were in contrast to the more subdued and natural tone of ChatBlackGPT. 

The passive sentiment observed in ChatGPT's response translated into its engagement with culturally nuanced questions, as exemplified by the AI assistant failing to use identity-specific characteristics in its outputs, even when explicitly mentioned by the users' prompt. While miniscule at glance, these microaggressions have the potential to contribute to user alienation \cite{winchester2023harmfulterms, antypas2024sensitive}. When identity was explicitly highlighted, ChatBlackGPT output responses specified user identity and were accompanied with concrete resources and affirming language. Of the resources that were given, ChatBlackGPT offered specific literature and notable figures for the user to research prior to beginning their travel journey. Given prior work supporting this idea that Black people lean on support groups for tangible resources \cite{park2022flourishing}, such artifacts are especially valuable.




Culturally-tailored AI assistants alone can not prevent the racial and discriminatory experiences that Black people experience, but they can serve as a resource to guide Black users who are otherwise underrepresented in accurately tailored information-seeking experiences. By highlighting the distinguishing features, and particularly the tailored resources that ChatBlackGPT generates in comparison to ChatGPT, our preliminary work emphasizes the need for further inquiries around \textit{how} Black people perceive AI assistants and specifically their use experiences with culturally tailored AI assistants. Minimal work in HCI has explored these perceptions, and virtually no work exists around perceptions and understanding of voice assistants when used for travel-related inquiries that impact Black communities. 


\subsection{Future Work: AI Assistants for the Black Lived Experience}
Based on our preliminary analysis as well as the lack of literature on Black people's perceptions and use of AI assistants more broadly ~\cite{slota2021methods, harrington2023trust, harrington2022s}, our future work aims to extend our understanding of what Black communities desire out of culturally tailored tools and evolve the current landscape of AI assistants to improve Black engagement. Therefore, we will deploy a survey for Black adults in the U.S. to gain a better understanding of their engagement with AI assistants more broadly. Findings from this survey will help ground our understanding of our target population's experiences and needs, which will impact the design of subsequent phases~\cite{mengesha2021don}. In order to understand users' past experiences using AI assistant tools, we will engage in semi-structured interviews to focus their challenges and desires out of AI assistants. We will explore themes related to perceptions of AI assistants and trust, with a focus on what use cases participants would envision themselves using AI assistants. Lastly, we will incorporate workshops to demo ChatBlackGPT among Black users to discuss their reactions and feedback as it relates to culturally-informed tools.

\section{Conclusion}

Through a comparative analysis of a culturally tailored versus a non-culturally tailored AI Assistant, we found that ChatBlackGPT offers concrete advice and culturally relevant resources when prompted about Black travel related inquiries. In contrast to the broader and less contextually situated responses offered by ChatGPT, ChatBlackGPT (as well as the creation of other forms of culturally tailored AI) can serve as a powerful resource for Black users looking to seek support for culturally nuanced inquiries. Our preliminary findings and proposed study design emphasize the need for further exploration around the perceptions and impacts of culturally tailored AI designed for Black communities in HCI.
\begin{acks}
We would like to thank the Google Academic Research Awards program for funding this work.
\end{acks}


\bibliographystyle{ACM-Reference-Format}
\bibliography{references}
\end{document}